\documentclass[conference]{IEEEtran}

\IEEEoverridecommandlockouts \IEEEpubid{\makebox[\columnwidth]{979-8-3503-21159/23/\$31.00~\copyright~2023~IEEE \hfill} \hspace{\columnsep}\makebox[\columnwidth]{ }} 

\IEEEoverridecommandlockouts
\usepackage{algorithm,algcompatible}

\usepackage{cite}
\usepackage{amsmath,amssymb,amsfonts}
\usepackage{graphicx}
\usepackage{textcomp}
\usepackage{xcolor}
\def\BibTeX{{\rm B\kern-.05em{\sc i\kern-.025em b}\kern-.08em
    T\kern-.1667em\lower.7ex\hbox{E}\kern-.125emX}}
\begin{document}

\title{Super-resolution of THz time-domain images \\based on low-rank representation\\
\thanks{This project has received funding from the European Union's Horizon 2020 research and innovation programme under grant agreement No. 101026453. This work was presented at the Sixth International Workshop on Mobile Terahertz Systems (IWMTS). %Copyright notice: 979-8-3503-2115-9/23/ \$31.00 ©2023 IEEE 
}
}

\author{\IEEEauthorblockN{1\textsuperscript{st} Marina Ljubenovi\'c}
\IEEEauthorblockA{\textit{Center for Cultural Heritage Technology} \\
\textit{Istituto Italiano di Tecnologia}\\
Venice, Italy\\
marina.ljubenovic@iit.it}
\and
\IEEEauthorblockN{2\textsuperscript{nd} Alessia Artesani}
\IEEEauthorblockA{\textit{Dep. of Biomedical Sciences} \\
\textit{Humanitas University}\\
%\textit{, Humanitas University, Milan, Italy}\\
Milan, Italy\\
alessia.artesani@hunimed.eu}
\and
\IEEEauthorblockN{3\textsuperscript{rd} Stefano Bonetti}
\IEEEauthorblockA{
\textit{Dep. of Molecular Sciences and Nanosystems} \\
\textit{University Ca' Foscari, Venice, Italy} \\
%Venice, Italy\\
\textit{Department of Physics}\\
\textit{Stockholm University, Sweden}\\
%Stockholm, Sweden\\
stefano.bonetti@unive.it}
\and
\IEEEauthorblockN{4\textsuperscript{th} Arianna Traviglia}
\IEEEauthorblockA{\textit{Center for Cultural Heritage Technology} \\
\textit{Istituto Italiano di Tecnologia}\\
Venice, Italy \\
arianna.traviglia@iit.it}
%\and
%\and
}

\maketitle

\begin{abstract}
Terahertz time-domain spectroscopy (THz-TDS) employs sub-picosecond pulses to probe dielectric properties of materials giving as a result a 3-dimensional hyperspectral data cube. 
% The ultra-broadband THz source, the frequency-dependent imaging process, and the point-like measurement procedure have tremendous effects on the spatial resolution of THz images. 
The spatial resolution of THz images is primarily limited by two sources: a non-zero THz beam waist and the acquisition step size. 
%THz images acquired with the smaller step size contain more details compared to ones acquired with bigger step sizes, but the acquisition time increases significantly. 
Acquisition with a small step size allows for the visualisation of smaller details in images at the expense of acquisition time, 
%(e.g. it takes three minutes to scan $5 \times 5$ mm area for 0.2 mm step size compared to 32 minutes for 0.1 mm step size)
but the frequency-dependent point-spread function remains the biggest bottleneck for THz imaging.
This work presents a super-resolution approach to restore THz time-domain images acquired with medium-to-big step sizes. %(e.g., 0.2 mm) and additionally corrupted by blur and noise. 
The results show the optimized and robust performance for different frequency bands (from 0.5 to 3.5 THz) obtaining higher resolution and additionally removing effects of blur at lower frequencies and noise at higher frequencies.
\end{abstract}


\section{Introduction}
%The terahertz (THz) region of electromagnetic field (between 0.1 and 10 THz) offers new imaging solutions for % applications in security \cite{2003_Kemp_Security}, 
%medicine \cite{2009_Pickwell_Medical}, conservation of cultural heritage \cite{2016_Cosentino_Cultural}, and in many other fields. 
THz time-domain spectroscopy (THz-TDS) employs sub-picosecond pulses to probe material properties of dielectric materials giving as a result a 3-dimensional hyperspectral (HS) data cube. This HS cube 
%which consists of several two-dimensional images corresponding to different frequency bands, 
might contain information on both surface and inner structures of the analysed sample. However, the ultra-broadband THz source (extending over more than 6 octaves), the frequency-dependent imaging process, and the point-like measurement procedure have tremendous effects on the spatial resolution of THz images. 

\begin{figure}[ht!]
\centering
\includegraphics[scale=0.09]{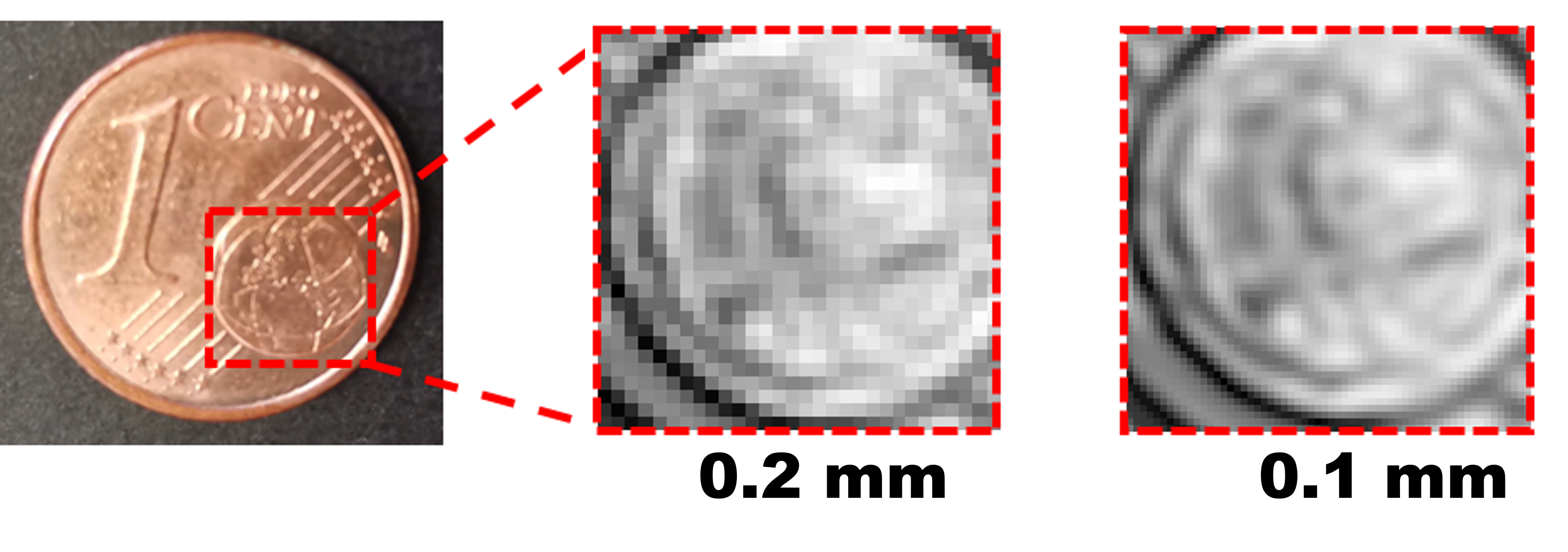}
\caption{Example of a THz HS band at 2.72 THz acquired with 0.2 and 0.1 mm step sizes.}
\label{fig:step_sizes}
\end{figure}

\noindent The spatial resolution of THz images is limited by two main sources: the point-spread function (PSF) of the imaging system and the acquisition step size.
A THz beam has a frequency-dependent divergence and it generates blurring degradation effects in HS images,
%The focal spot of a THz beam, with its complex structure, determines frequency-dependent blurring degradation effects, 
i.e., each frequency band is blurred with a different PSF, accompanied by the system-induced noise. The analysis of a surface is obtained in the reflection or transmission mode by raster-scanning the sample with a predefined scanning step size. 
%(for the THz system used in this work, step sizes in both scanning directions can be fixed equal to 0.1, 0.2, 0.5, 1 or 2 mm). 
An example of the amplitude signal of a raster-scanned sample taken with two different step sizes is shown in Fig. \ref{fig:step_sizes}. THz images acquired with smaller step sizes (e.g., 0.1 mm) contain more details compared to ones acquired with a medium step size (e.g., 0.2 mm), but the acquisition time increases significantly. For instance, it takes 3 minutes to scan $5 \times 5$ mm area for a 0.2 mm step size compared to 32 minutes for a 0.1 mm step size.

\noindent Different model-based and machine learning-based methods have been proposed for THz image resolution enhancement, mostly focused on removing the blurring effects introduced by the system PSF. 
Initial model-based methods tried to use simulated \cite{2017_Ahi_Mathematical,2019_Wong_Computational} or measured \cite{2008_Li_SuperRes,2010_Popescu_Point} PSF of the THz imaging system in combination with well-known single image deblurring approaches such as Richardson-Lucy \cite{2014_Xu_High-Resolution,2019_Wong_Computational} and Wiener filtering \cite{2010_Popescu_Point}. %Recently, a method based on physical properties of a THz beam propagation tailored to biological samples is proposed in \cite{2021_Freer_Hybrid}.
Other methods exploited state-of-the-art deblurring approaches based on \textit{total variation} (TV) and/or sparsity \cite{2019_Wong_Computational}. All these methods suffer from several limitations: i) they are focused on single images and not easily applied to HS images; ii) they often amplify noise present at high-frequency bands; and iii) they provide images of the same size as input images, thus performing only deblurring (i.e., removing of PSF effects) and not a super-resolution procedure (i.e., decreasing of pixel size). 
%Recently, a deblurring method that exploits a low-rank property of THz HS images was proposed \cite{2020_Ljubenovic_Joint}. This method reduces the effects of blur and noise, but does not increase a resolution of THz HS images. 

\noindent Machine learning-based image deblurring and super-resolution approaches applied to RGB images provide state-of-the-art results \cite{2014_Xu_DeepDeconvolution}. Recently, several papers proposed the implementation of such approaches to THz images in order to exploit the powerful mapping ability of the convolution neural network (CNN) and the efficient training GPU implementation \cite{2020_Ljubenovic_THzCNN,2020_Li_Adaptive,2021_Wang_THz_SR_CCNN}. Other methods utilize different types of NNs such as a local-pixel graph neural network \cite{2021_Lei_LocalPixel} and a residual generative adversarial network \cite{2023_Hou_ResidualTHz}.
The majority of these methods are trained on synthetic images or images that belong to a single class such as biological images \cite{2021_Lei_LocalPixel}, thus providing limited results when tested on real data. Obtaining a high enough number of real data for building training datasets is time-consuming and for many applications impractical. Moreover, most CNN-based methods are tailored to a single-frequency THz image and do not take into consideration the continuous change of PSF with frequency present in THz HS images. 

\noindent Methods that use computer vision techniques for performing HS image deblurring and super-resolution were mostly developed for remote sensing applications relaying to dimensionality reduction approaches \cite{2013_Liao_HSIDeblurring,2018_Dian_HSISuperResolution, 2021_Ljubenovic_Improved}. 
%HS image deblurring methods are often based on low-rank property of HS images exploiting dimensionality reduction approaches such as \textit{principal component analysis} (PCA) \cite{2013_Liao_HSIDeblurring} or \textit{singular value decomposition} (SVD) \cite{2021_Ljubenovic_Improved}. 
Super-resolution methods for HS images are usually based on data fusion: a low-resolution image (or bands) is fused with a high-resolution image (e.g., a panchromatic image) or higher resolution bands to obtain a high-resolution HS or multispectral image  \cite{2015_Simoes_ConvexSR,2017_Lanaras_SRMultispectral}. 
%Similarly, information from high-resolution bands can be utilized to increase the resolution of low-resolution bands of multiresolution MS images (e.g. Sentinel-2, MODIS)\cite{2017_Lanaras_SRMultispectral,2019_Paris_Novel}. 
%With applying dimensionality reduction, the data are represented in a lower-dimensional subspace, reducing the dimensionality of the problem and significantly improving its conditioning and thus, the influence of noise. 
Although these methods give promising results, they require additional input data, such as a high-resolution image of the same sample, that are not available when working with THz images. Additionally, they are tailored to blur degradation introduced by a hyperspectral remote sensing camera that is often assumed to be uniform over bands.  

%\noindent\textbf{Contribution.} 
Recently, a deblurring and denoising method that jointly reduces the blur and noise effects, without increasing the resolution of the THz HS images was developed for images acquired in the transmission mode \cite{2020_Ljubenovic_Joint}. A similar approach was tested on THz images in the 0.25–6 THz range, acquired in the reflection mode \cite{2022_Ljubenovic_BeamShape}. Here, the methodology is improved by including a super-resolution approach to restore THz HS images acquired with bigger step sizes, additionally corrupted by blur and noise. By restoring low-resolution THz images digitally, the acquisition time is significantly reduced. 
The proposed approach is inspired by state-of-the-art remote sensing methods based on dimensionality reduction \cite{2017_Lanaras_SRMultispectral} with the implementation of an important step to accommodate frequency-dependent PSFs. It relies on the assumption that spectral bands of THz time-domain images are correlated and can thus be represented in a lower-dimensional subspace, where most of the useful information is contained. The subspace is learned from the input data and thus, the method does not require additional parameters. Moreover, the super-resolution problem is formulated as a minimisation of a convex objective function with an edge-preserving regularizer. This minimisation is computed by an efficient numerical solver based on the alternating direction method of multipliers (ADMM) algorithm. 
In this way, the resolution of the digital image is increased, intended as pixel density, and at the same time, the noise and the blur effects that corrupt the broadband images (from 0.25 to 3.5 THz) are reduced. The super-resolution process is accomplished in accordance with the frequency band of the image and the additional pixels are created with an iterative method that considers the PSF of the system. Here, only super-resolution that doubles the number of pixels in both directions is considered (e.g., an image with $100 \times 100$ pixels is increased to an image with $200 \times 200$ pixels) as in this case, newly created pixels are formed by information coming from four closest neighboring pixels. 
% To the best of our knowledge, this is the first time that a super-resolution approach based on the low-rank property of HS images is proposed to restore THz time-domain images. 

%===========================================================================
\section{Low-rank representation of hyperspectral data}
\label{problem}

If $B$ and $n_L$ denote the number of bands and the number of low-resolution pixels, respectively, then vectorised observed image can be represented as $\textbf{y} = (\textbf{y}_1,\textbf{y}_2,...,\textbf{y}_B) \in \mathbb{R}^{Bn_L}$, where $\textbf{y}_i$, for $i = 1,...,B$, stands for the pixel intensities of each individual band collected into a vector. Given the upsampling factor $d$, the number of low-resolution pixel is calculated as $n_L = \frac{n}{d^2}$ with $n$ representing the (desired) number of high-resolution pixels. Thus, the underlying high-resolution vectorised image is $\textbf{x} = (\textbf{x}_1,\textbf{x}_2,...,\textbf{x}_B) \in \mathbb{R}^{Bn}$. In the matrix form, the output image can be reformatted to $\textbf{X} = [\textbf{x}_1^T; \textbf{x}_2^T;...;\textbf{x}_B^T] \in \mathbb{R}^{B \times n}$, leading to $\textbf{x} = \text{vec}(\textbf{X}^T)$. The observation model with the assumption of Gaussian noise, $\textbf{n}$, is represented as
\begin{equation}
\label{equ:obs_model}
\textbf{y} = \textbf{M}\textbf{H}\textbf{x} + \textbf{n},
\end{equation}
where $\textbf{M} \in \mathbb{R}^{Bn_L \times Bn}$ and $\textbf{H} \in \mathbb{R}^{Bn \times Bn}$ represent the sampling matrix (i.e., sampling of $\textbf{x}$ to obtain $\textbf{y}$) and a 2D cyclic convolution associated with the PSF of the corresponding band at the highest spatial resolution, respectively. The blur matrix $\textbf{H} = \text{bkdiag}(\textbf{H}_1,...,\textbf{H}_B)$ is a block-circulant-circulant-block (BCCB) matrix depicting a different blur for each frequency band. The model in (\ref{equ:obs_model}) has fewer observations than unknowns and thus, is ill-posed. 

\noindent Considering the high correlation between bands, we assume that the columns of $\textbf{X}$ (i.e., spectral vectors) live in a lower-dimensional subspace $\mathcal{S}_p$, with $p \ll B$ and thus can be represented as 
\begin{equation}
\label{equ:subspace}
\textbf{X} = \textbf{E}\textbf{Z},
\end{equation}
where $\textbf{E} = [\textbf{e}_1,...,\textbf{e}_p] \in \mathbb{R}^{B \times p}$ stands for basis of $\mathcal{S}_p$ and  $\textbf{Z} \in \mathbb{R}^{p \times n}$ holds the representation coefficients of $\textbf{X}$ in $\mathcal{S}_p$.
By assuming that $\textbf{E}$ is semi-unitary, vectorisation of $\textbf{X}$ leads to $\textbf{x} = (\textbf{E} \otimes \textbf{I})\textbf{z}$, with $\textbf{I}$ as an identity matrix. With the dimensionality reduction, the observation model becomes
\begin{equation}
\label{equ:obs_model_z}
\textbf{y} = \textbf{M}\textbf{H}(\textbf{E} \otimes \textbf{I})\textbf{z} + \textbf{n}.
\end{equation}
If $pn < Bn_L$, the problem is no longer ill-posed. However, due to the assumption of the cyclic convolution, it is still ill-conditioned and thus, sensitive to the presence of (even low-level) noise in the observed image. 
%===========================================================================

\subsection{THz Beam Profile Estimation}

Each frequency band of THz time-domain images is blurred with a different PSF.  The minimum beam
radius (beam waist) of a THz beam is wider at lower frequencies resulting in blurrier bands. On the contrary, at higher frequencies, the beam waist is smaller and thus bands are sharper. However, due to lower amplitudes at higher frequencies, these bands are noisier.

\noindent In this work, the intensity profile of a THz beam is assumed to have Gaussian distribution with the beam intensity calculated as
\begin{equation}
\label{equ_intensity}
I(r,z) = \frac{2I_p}{\pi w^2(z)}\text{exp} \left(\frac{-2r^2}{w^2(z)}\right),
\end{equation}
where $I_p$ is the beam power, $w(z)$ is the beam radius in the direction $z$, and $r$ is the distance from the beam axis. The radius of the beam varies along the propagating direction as
\begin{equation}
\label{equ_radius}
w(z) = w_0 \sqrt{1 + \left(\frac{z}{\pi w_0^2/\lambda}\right)^2},
\end{equation}
where  $w_0 = \frac{4}{\pi} \lambda \frac{f_L}{D}$ represents the half of the beam waist and $\lambda$ the beam wavelength (the beam waist $2w_0$ depends on the wavelength $\lambda$ correlated to band frequencies). $f_L$ and $D$ represent the focal length and diameter of the focusing THz lens. PSF of each band corresponds to an intersection of the THz beam calculated by using (\ref{equ_intensity}) and (\ref{equ_radius}) with an orthogonal plane. 
%===========================================================================

\section{Frequency-Optimized Super-Resolution}
\label{method}

To obtain a high-resolution image, we solve the optimization problem
\begin{equation}
\label{equ:optim_problem}
\hat{\textbf{z}} \in \underset{\textbf{z}}{\mathrm{argmin}} ||\textbf{M}\textbf{H}(\textbf{E} \otimes \textbf{I})\textbf{z} - \textbf{y}||^2 + \gamma \Phi_{\textbf{w},\textbf{c}}(\textbf{z}), 
\end{equation}
where $\Phi_{\textbf{w},\textbf{c}}(\textbf{z})= \sum_{i=1}^p \sum_{j=1}^n (c_iw_j(\textbf{F}_h\textbf{z}_i)_j^2 + c_iw_j(\textbf{F}_v\textbf{z}_i)_j^2)$ is a regularization term with weights $\textbf{w}$ and $\textbf{c}$, and the regularization parameter $\gamma$. $\textbf{D}_h = \textbf{I} \otimes \textbf{F}_h$ and $\textbf{D}_v = \textbf{I} \otimes \textbf{F}_v$ represent approximate horizontal and vertical derivatives of $\textbf{z}$.

\noindent The optimization problem is tackled by applying the instance of the ADMM algorithm, SALSA, \cite{2010_Afonso_Fast} by transforming the unconstrained problem from (\ref{equ:optim_problem}) to a constrained one, as follows:
%\begin{equation}
%\begin{aligned}
%\label{equ:admm_constrained}
%&\underset{\textbf{z}, \textbf{v}_1, \textbf{v}_2, \textbf{v}_3}{\text{min}} \hspace{2mm} 
%&&||\textbf{M}\textbf{H}\textbf{v}_1 - \textbf{y}||^2 + \gamma %\Phi_{\textbf{w},\textbf{c}}(\textbf{v}_2,\textbf{v}_3)\\
%&  \text{subject to} 
%&& \textbf{v}_1 = (\textbf{E} \otimes \textbf{I})\textbf{z} \\
%&
%&& \textbf{v}_2 = \textbf{D}_h\textbf{z}\\
%&
%&& \textbf{v}_3 = \textbf{D}_v\textbf{z}.
%\end{aligned}
%\end{equation}

\begin{equation}
\begin{aligned}
\label{equ:admm_constrained}
&\underset{\textbf{z}, \textbf{v}_1, \textbf{v}_2, \textbf{v}_3}{\text{min}} \hspace{2mm} 
&&||\textbf{M}\textbf{H}\textbf{v}_1 - \textbf{y}||^2 + \gamma \Phi_{\textbf{w},\textbf{c}}(\textbf{v}_2,\textbf{v}_3)\\
&  \text{subject to} 
&& \textbf{v}_1 = (\textbf{E} \otimes \textbf{I})\textbf{z}; \hspace{2mm} \textbf{v}_2 = \textbf{D}_h\textbf{z};
\hspace{2mm} \textbf{v}_3 = \textbf{D}_v\textbf{z}.
\end{aligned}
\end{equation}
The Augmented Lagrangian of the above problem using the vectors of Lagrange multipliers $\textbf{d}_1$, $\textbf{d}_2$, and $\textbf{d}_3$ is
\begin{equation}
\begin{aligned}
\label{equ:lagrangian}
\mathcal{L}(\textbf{z}, \textbf{v}_1, \textbf{v}_2, \textbf{v}_3, \textbf{d}_1, \textbf{d}_2, \textbf{d}_3) = 
||\textbf{M}\textbf{H}\textbf{v}_1 - \textbf{y}||^2 \\
+ \frac{\mu_1}{2}||(\textbf{E} \otimes \textbf{I})\textbf{z} - \textbf{v}_1 - \textbf{d}_1||^2 + 
\gamma \Phi_{\textbf{w},\textbf{c}}(\textbf{v}_2,\textbf{v}_3) \\
+ \frac{\mu_2}{2}||\textbf{D}_h\textbf{z} - \textbf{v}_2 - \textbf{d}_2||^2 + 
\frac{\mu_3}{2}||\textbf{D}_v\textbf{z} - \textbf{v}_3 - \textbf{d}_3||^2,
\end{aligned}
\end{equation}
with the penalty parameters $\mu_i \geq 0$ for $i = 1,2,3$. The ADMM addresses the problem by alternatingly minimizing (\ref{equ:lagrangian}) over $\textbf{z}$, $\textbf{v}_1$, $\textbf{v}_2$, and $\textbf{v}_3$ and by updating the vector of Lagrange multipliers $\textbf{d}_1$, $\textbf{d}_2$, and $\textbf{d}_3$ by keeping the other variables fixed. The proposed method named Frequency-Optimized Super-Resolution (FO-SupRes) is presented in Algorithm 1. 

%%%Algorithm ADMM
\begin{algorithm}
\small
\label{alg_admm}
    \caption{FO-SupRes}
  \begin{algorithmic}[1]
    \STATE \textbf{Input:} Observed image $\textbf{y}$, regularization parameter $\gamma$, THz system frequencies ($\text{Freq}$), weights $\textbf{w}$ and $\textbf{c}$
    \STATE \textbf{Initialization}: Set $k = 0$; Initialize $\textbf{v}_1^{(0)}$, $\textbf{v}_2^{(0)}$, $\textbf{v}_3^{(0)}$, $\textbf{d}_1^{(0)}$, $\textbf{d}_2^{(0)}$, and $\textbf{d}_3^{(0)}$
    \WHILE{stopping criterion is not satisfied}
      \STATE $k \leftarrow k + 1$
      \STATE Calculate PSFs for each band using (\ref{equ_intensity}) and (\ref{equ_radius}) in the focus of the beam (minimum waist) 
      \STATE Minimize $\textbf{z}^{(k)}$ by keeping other variables fixed
 	  \STATE Minimize $\textbf{v}_1^{(k)}$ by keeping other variables fixed
	  \STATE Minimize $\textbf{v}_2^{(k)}$ by keeping other variables fixed
	  \STATE Minimize $\textbf{v}_3^{(k)}$ by keeping other variables fixed
	  \STATE Update $\textbf{d}_1^{(k)}$, $\textbf{d}_2^{(k)}$, and $\textbf{d}_3^{(k)}$
    \ENDWHILE
  \end{algorithmic}
\end{algorithm}
%=================================================================================================

\section{Experimental results}
\label{results}

In all the experiments, the following settings are used for the ADMM algorithm: the number of iterations is set to 100, $\mu_i = 0.2$ and $\textbf{d}_i^{(0)} = 0$, for $i = \{1,2,3\}$, and the image estimate is initialised with the estimate from the previous iteration. The number of subspaces is set to 10, the sampling factor is set to $d = 2$, and the parameter $\gamma$ is hand-tuned for the best visual results.
Results obtained with the proposed super-resolution algorithm, named Frequency-Optimized Super-Resolution (FO-SupRes), are compared with the results of the conventional bicubic interpolation (Bicubic) and a state-of-the-art CNN-based super-resolution algorithm (SRCNN)\cite{2015_Dong_SRCNN}. Without access to a training dataset, it was not possible to fairly test NN-based super-resolution methods for single-band THz images. %\cite{2020_Li_Adaptive}. 

\begin{figure}[ht!]
\centering
\includegraphics[scale=0.15]{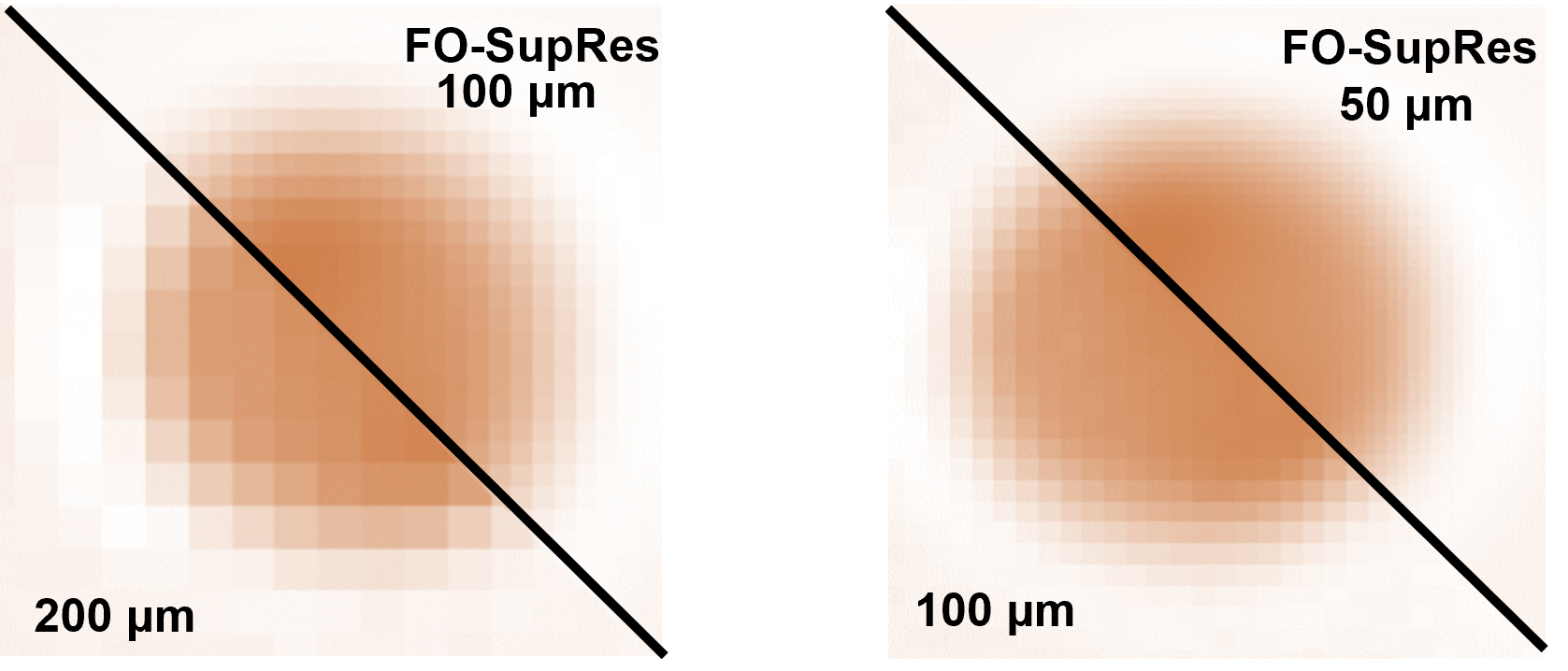}
\caption{Examples of raw and estimated bands at 1.5 THz acquired with two different step sizes (0.2 and 0.1 mm).}
\label{fig:holes_proposed}
\end{figure}

\noindent Experiments are performed in the reflection mode using two samples: a hole on a metallic plate and 1 cent coin. Fig. \ref{fig:holes_proposed} illustrates the results of the proposed super-resolution algorithm with two different step sizes (0.2 and 0.1 mm) obtained on the simple sample, a hole on a metallic plate. 

\begin{figure}[ht!]
\centering
\includegraphics[scale=0.14]{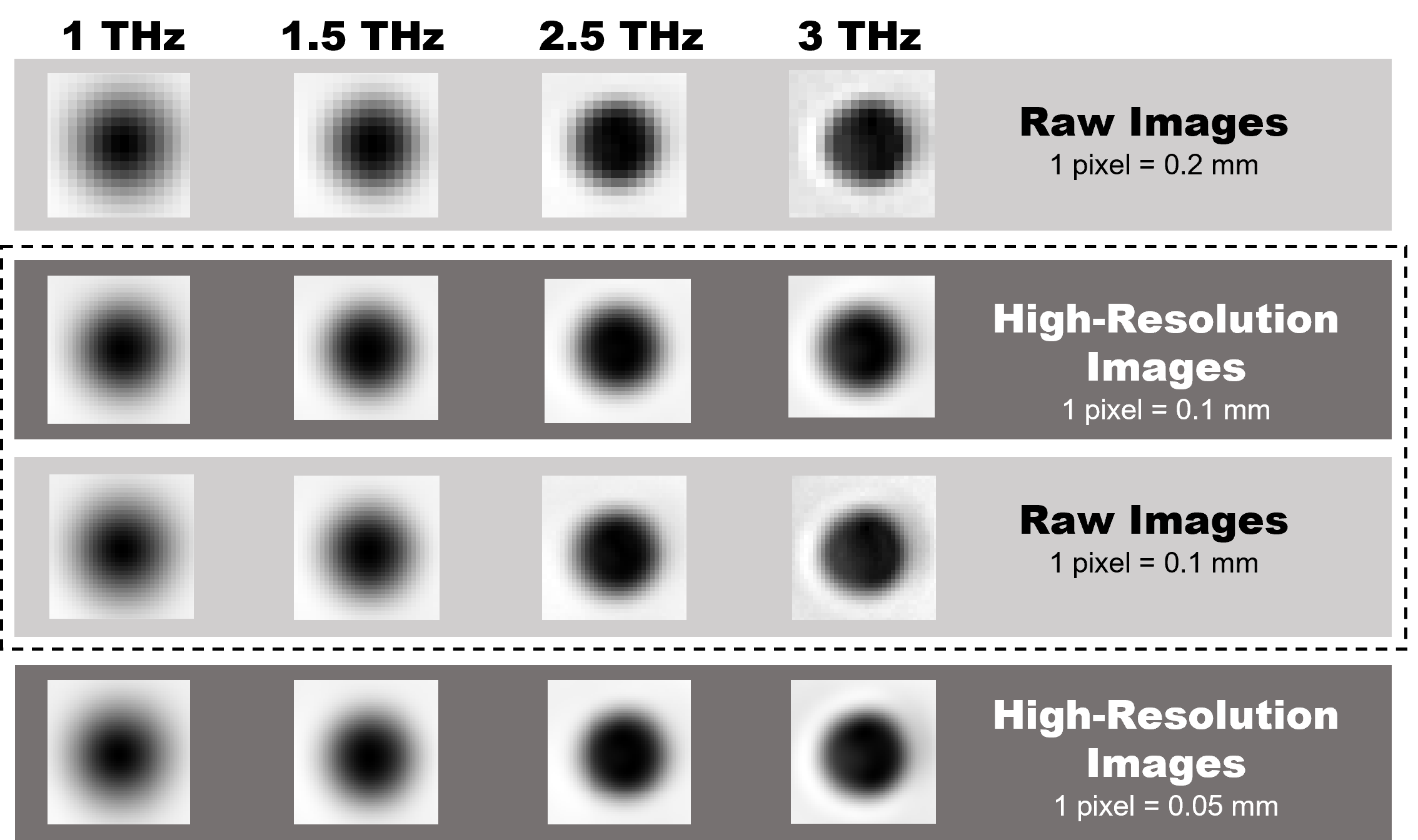}
\caption{Results obtained on a hole on a metallic place; Rows from up to bottom: Raw bands (0.2 mm), restored high-resolution bands with FO-SupRes (0.1 mm), raw bands (0.1 mm), and restored high-resolution bands with FO-SupRes (0.05 mm).}
\label{fig:holes_compare}
\end{figure}

\noindent Fig. \ref{fig:holes_compare} shows estimated bands corresponding to four frequencies (i.e., 1, 1.5, 2.5, and 3 THz) acquired by two different step sizes (i.e., 0.2 and 0.1 mm). Medium frequencies of the THz range (e.g., from 1 to 3 THz) are not corrupted by severe noise and blur and thus, the influence of the step size choice is mostly visible in the bands selected from that range. Results show that the performance of the proposed approach is stable over bands, i.e., blur and noise are successfully removed and the resolution increased. Additionally, images digitally restored to achieve 0.1 mm resolution, i.e., 1 pixel = 0.1 mm (second raw), are visually comparable with the raw images acquired with the same resolution (third row).

\begin{figure}[ht!]
\centering
\includegraphics[scale=0.137]{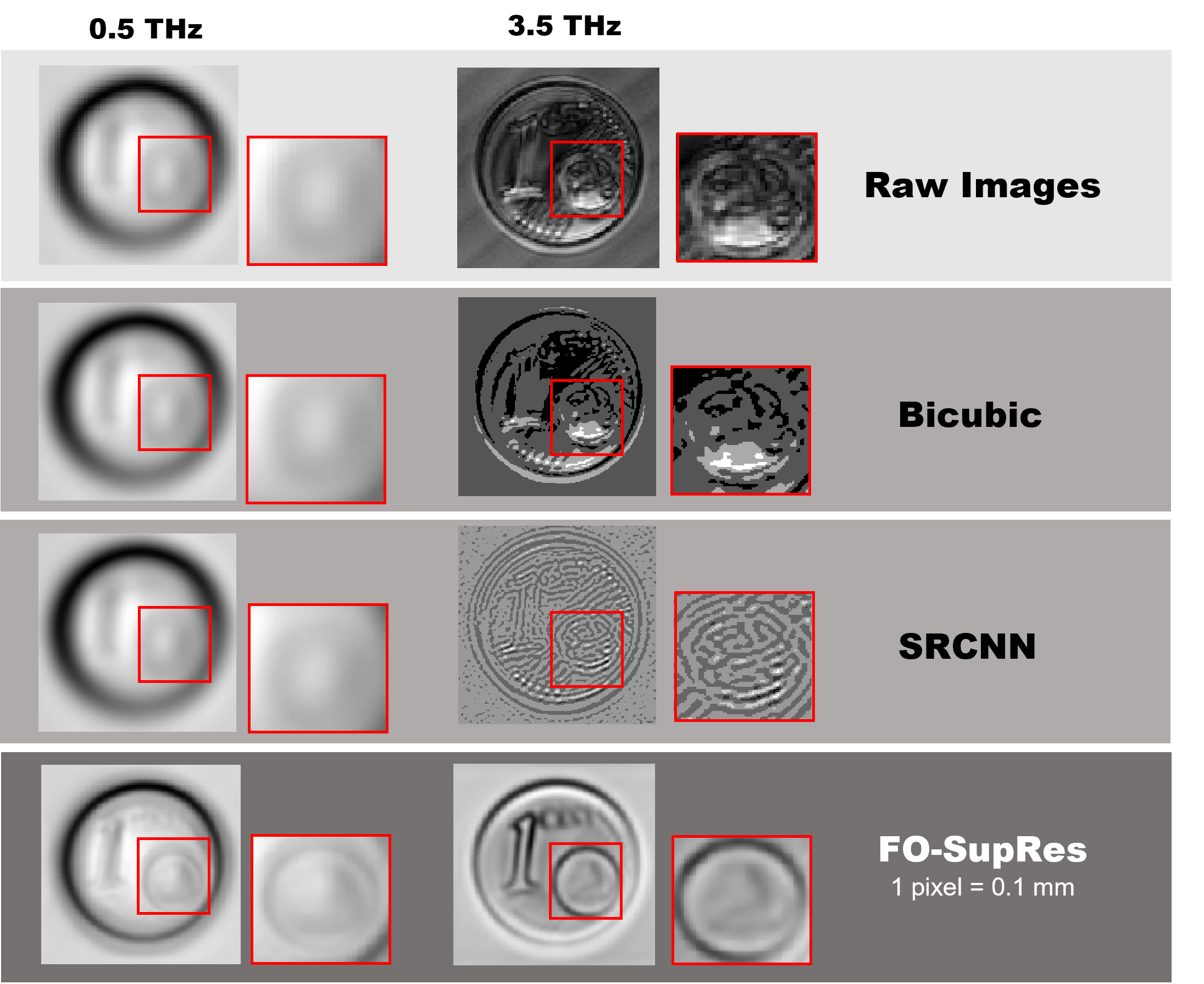}
\caption{Results obtained on 1 cent coin. Bands presented (from up to bottom): Raw bands, estimated bands with bicubic interpolation, estimated bands with SRCNN, and estimated bands with the proposed approach.}
\label{fig:cent_compare}
\end{figure}

\noindent Figure \ref{fig:cent_compare} shows the results obtained on 1 cent coin acquired with 0.2 mm step with the proposed approach, FO-SupRes, compared to the results obtained by conventional bicubic interpolation and state-of-the-art neural network-based approach (SRCNN). Here, only one band corresponding to a low frequency (0.5 THz) and a band corresponding to a medium-high frequency are presented for the sake of clarity. The proposed method is optimized over all frequency bands leading to a reduction in both blur and noise and an increase in resolution. Contrary to that, bicubic interpolation and SRCNN fail to remove blur from bands corresponding to lower frequencies and show significant artefacts on bands corresponding to higher frequencies. These artefacts are most likely caused by the presence of (even a small) amount of noise as both methods are tailored to noiseless images: bicubic interpolation mistakes noise with a valid data point during calculation and the SRCNN method is trained only on noiseless images.   

\begin{figure}[ht!]
\centering
\includegraphics[scale=0.17]{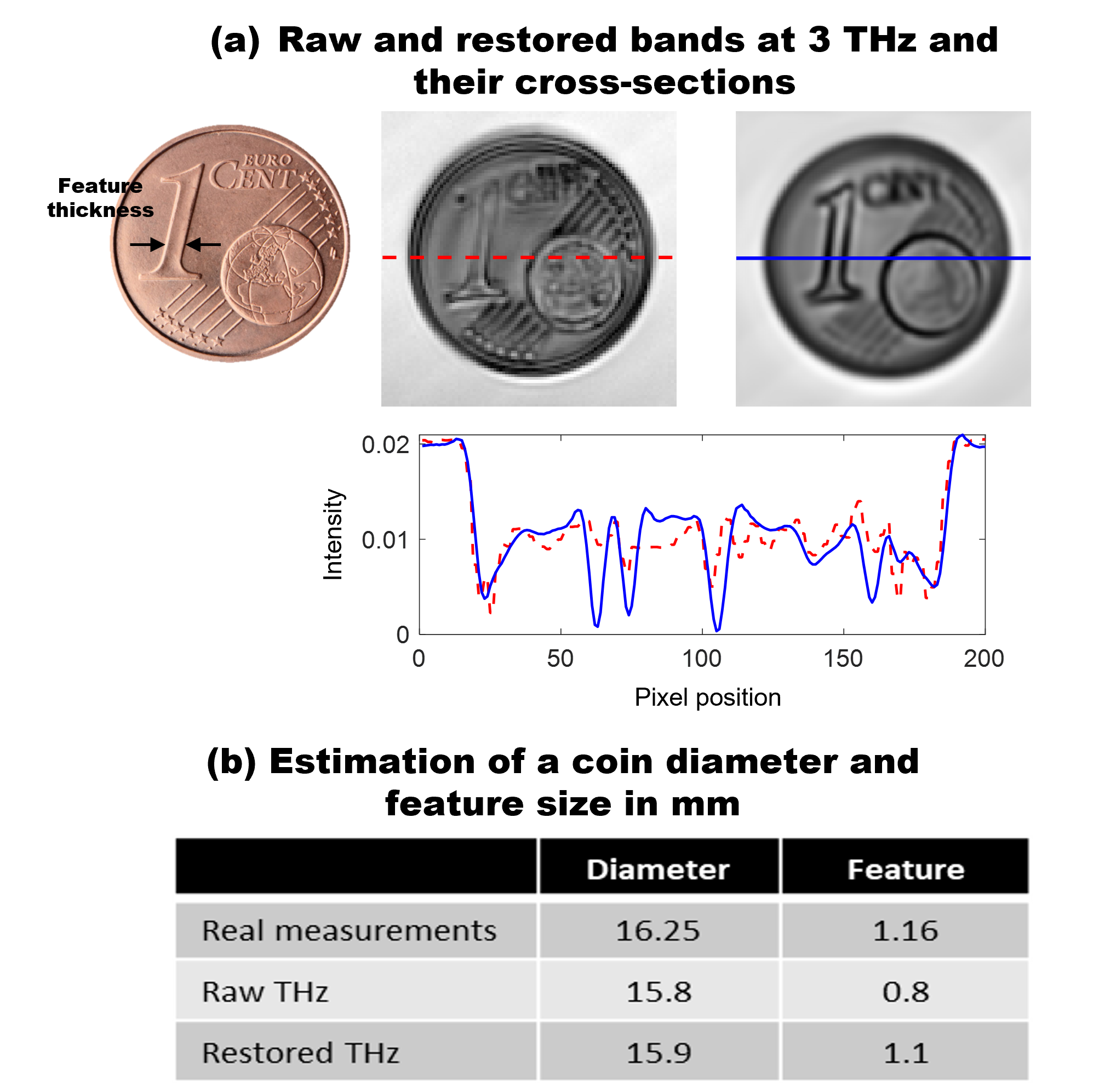}
\caption{Measuring a feature thickness: (a) Cross-sections of raw and restored bands and (b) estimated feature thickness and the sample diameter.}
\label{fig:feature}
\end{figure}

\noindent The influence of the proposed approach on the sharpness of a feature is presented in Fig. \ref{fig:feature} visually through cross-sections (i.e., a cross-section of a raw band is compared to a cross-section of a restored band) and in terms of feature size and a coin diameter estimation. Results show that to achieve an estimation that is closer to real values, super-resolved images should be used as a pre-processing step. 

%===========================================================================

\section{Conclusion}
\label{conclusion}

This paper has demonstrated the results of a super-resolution approach tailored to THz time-domain images that made it possible to jointly restore all the bands of an HS cube by exploiting the low-rank property of HS data and an edge-preserving regularizer. 
To the best of our knowledge, this is the first time that a super-resolution approach based on the low-rank property of HS images is proposed to restore THz time-domain images. 
As a result of such an approach, images acquired with the medium step size (e.g., 0.2 mm) and shorter acquisition time were digitally restored and achieved a resolution similar to that of images acquired by a smaller step size (e.g., 0.1 mm). The results show the optimized and robust performance for different frequency bands (from 0.5 to 3.5 THz) obtaining higher resolution and additionally removing effects of blur at lower frequencies and noise at higher frequencies, without introducing new artefacts. The proposed approach may have a tremendous impact on applications for which a high resolution is crucial, but the acquisition time is limited (e.g., in-line inspection to identify corrosion, cracks, and other defects). 
%The future work will include the automatic estimation of PSFs corresponding to each band before performing super-resolution.

%======================================================================
% Bibliography
\bibliographystyle{./IEEEtran}
\bibliography{./IEEEabrv,./THz_SR}

\end{document}